\documentclass[fleqn,usenatbib]{mnras}

\usepackage{newtxtext,newtxmath}
\usepackage[T1]{fontenc}
\usepackage{ae,aecompl}

\usepackage{graphicx}	% Including figure files
\usepackage{amsmath}	% Advanced maths commands
\usepackage{amssymb}	% Extra maths symbols

\title[NY Vir Synchronization]{Convection physics and tidal synchronization of the subdwarf binary NY Virginis}

\author[H. P. Preece et al.]{
Holly P. Preece,$^{1,2}$\thanks{E-mail: hpp25@ast.cam.ac.uk}
Christopher. A. Tout,$^{1}$
C. Simon. Jeffery$^{2}$
\\
$^{1}$Institute of Astronomy, University of Cambridge, Cambridge\\
$^{2}$Armagh Observatory and Planetarium, Armagh, BT61 9DG \\
}

\date{Accepted XXX. Received YYY; in original form ZZZ}

\pubyear{2018}

\begin{document}
\label{firstpage}
\pagerange{\pageref{firstpage}--\pageref{lastpage}}
\maketitle

\begin{abstract}
Asteroseismological analysis of NY Vir suggests that at least the outer 55 per cent of the star (in radius) rotates as a solid body and is tidally synchronized to the orbit. Detailed calculation of tidal dissipation rates in NY Vir fails to account for this synchronization. Recent observations of He core burning stars suggest that the extent of the convective core may be substantially larger than that predicted with theoretical models. We conduct a parametric investigation of sdB models generated with the Cambridge STARS code to artificially extend the radial extent of the convective core. These models with extended cores still fail to account for the synchronization. Tidal synchronization may be achievable with a non-MLT treatment of convection. 

\end{abstract}

\begin{keywords}
stars: subdwarfs -- stars: binaries: close -- stars: interiors -- stars: rotation -- stars: horizontal branch
\end{keywords}

%%%%%%%%%%%%%%%%%%%%%%%%%%%%%%%%%%%%%%%%%%%%%%%%%%

%%%%%%%%%%%%%%%%% BODY OF PAPER %%%%%%%%%%%%%%%%%%

\section{Introduction}

Hot subdwarf B (sdB) stars are core-helium burning stars which have had their hydrogen-rich envelopes stripped, most likely in a binary interaction. The stars are typically slow rotators. However, those in close binaries are somewhat spun up. The sdB stars in close binaries, with orbital periods less than $10\,\rm{d}$, have either low-mass main-sequence or white dwarf companions. The companions are unseen so it is not possible to measure the inclination of the observed systems unless they are eclipsing. If the system is tidally locked then the spin period and orbital period of the binaries should be the same and an observed rotation velocity would allow the inclination to be measured. Several observed sdB systems challenge this assumption \citep{pablo2nonsync,pablob4,j1622+4730}.  Theoretical calculations of tidal synchronization time-scales for sdB stars fail to account for  synchronization via either the equilibrium or dynamical dissipation mechanisms \citep{preece}.

Of the observed pulsating sdB binaries, the eclipsing HW\,Vir type binary NY\,Vir (PG\,1336$-$018) is the only object whose outer layers show evidence of synchronous rotation with the binary orbit \citep{charpinetnyvir}. The star oscillates with p-modes in its outer 55 per cent. Rotation in the deep interior is not constrained owing to the lack of sensitivity of p-modes to these regions.

\cite{charpinetnyvir} obtained a mass for the sdB component of $0.459\pm 0.006\,\rm{M_\odot}$ from asteroseismology, while  \citet{nyvirmass13} measured an asteroseismic mass of $0.471\,\pm \, 0.006\,\rm{M_\odot}$. \cite {nyvirsyncmaja} obtained three equally probable solutions from photometry and radial velocities. These give sdB masses of $0.530$, $0.466$ or $0.389\,\rm{M_\odot}$. The companion mass $M_2$ is either $0.11$ or $0.12\,\rm{M_\odot}$ and the orbital period of the binary $P_{\rm{orb}} = 0.101016\, \rm{d}$.

\begin{figure}
	\includegraphics[width=\columnwidth]{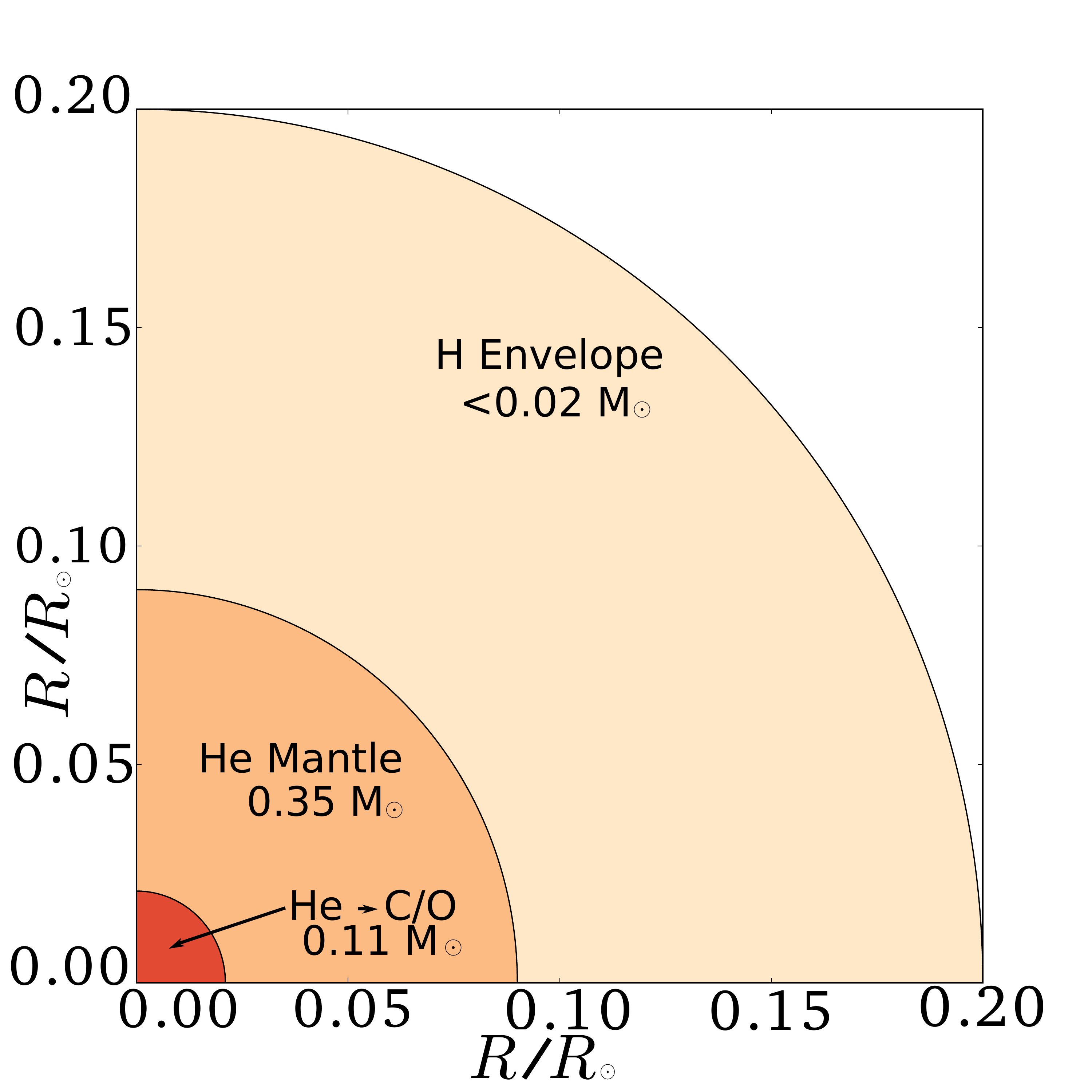}
	\caption{Scale diagram of a typical sdB star as predicted by stellar models. He $\to$ C/O (red) indicates the convective zone.}
	\label{fig:sdbdiag}
\end{figure}

The tidal synchronization time-scale is inversely proportional to the ratio of the radius of the dissipative region to the binary separation to the sixth power \citep{darwin,eggzbook}. Increasing the radius of the convective region reduces the tidal synchronization time. Observational asteroseismic data suggest the radial extent of the He burning core, as illustrated in Fig. \ref{fig:sdbdiag}, is substantially underestimated in stellar models \citep{vangrootel2010a,vangrootel2010b,charpinet2011,naturecores}. We investigate whether increasing the radius of the convective zone could reduce synchronization times sufficiently to account for the observed synchronization of NY\,Vir. We examine the effect that increasing the extent of the convective region has on all the quantities which go into the tidal synchronization calculations.

\section{Stellar Models}
The evolutionary models used in this study were all constructed with the Cambridge STARS code as first described by \cite{eggz71} and subsequently updated by \cite{chrisupdate} and \cite{stancliffe09}. Three classes of model were created, one with overshoot (labelled: $\delta_{\rm ov}=0.12$), one without overshoot ($\delta_{\rm ov}=0$) and one without overshooting but with a modified Schwarzschild criterion ($\Delta \nabla + 0.15$), where $\Delta \nabla \equiv \nabla_{\rm{r}} - \nabla_{\rm{a}}$, the difference between radiative and adiabatic thermodynamic gradients $\rm{d}\,ln\,\it{T} / \rm{d}\,ln\,\it{P}$. Under the standard Schwarzschild criterion, the convective region is defined as that where $\Delta \nabla > 0$ and hence includes the semiconvective region. The super-adiabacity of the convective region that develops as the star evolves is very low and is in fact more likely a semi-convective region. For each of these classes, an early and a late model were compared. The early model was defined as the model obtained when the fractional core He abundance by mass drops to 0.9. The late model was defined to be the model where the convective core reached its maximal radial extent. The initial models were constructed without a modified Schwarzschild criterion by the same method used by \cite{preece}. We introduce several mechanisms for artificially increasing the convective region.

\subsection{Modifying the Schwarzschild Criterion}
For the models labelled $\Delta \nabla + 0.15$, the extent of the convective region was artificially extended by modifying the Schwarzschild criterion for stability against convection from $\nabla_{\rm{r}} - \nabla_{\rm{a}} > 0$  to $\nabla_{\rm{r}} - \nabla_{\rm{a}} + 0.15 > 0$. This has the effect of forcing convection to occur in regions near to convective boundaries which would otherwise be radiative. The increment 0.15 was chosen because this was the largest which produced stable evolutionary models.

\subsection{Semi-convection and Overshooting}
\cite{starssemiconv} implemented semiconvection in STARS as a diffusive process which follows \cite{schwarzscsemiconv}'s prescription. It assumes that the energy transport by convection in the semiconvective region is borderline negligible but that there is substantial chemical mixing which avoids any discontinuity in the chemical profile.  Semiconvective regions then have $\nabla_{\rm{r}} \approx \nabla_{\rm{a}}$.

For convective overshooting we introduce a parameter $\delta$ such that convection occurs when 
\begin{equation}
\nabla_{\rm{r}} - \nabla_{\rm{a}} >  - \delta,
\end{equation}
where $\nabla_{\rm{r}}$ and $\nabla_{\rm{a}}$ correspond to the radiative and adiabatic thermodynamic gradients $\partial \ln \it{T} / \partial \ln \it{P}$, respectively, and where the overshooting parameter $\delta$ is
\begin{equation}
\delta = \frac{\delta_{\rm{ov}}}{2.5+20\zeta + 16 \zeta^2}.
\end{equation}
Here $\zeta$ is the ratio of radiation pressure to gas pressure and $\delta_{\rm{ov}}$ is a user defined parameter calibrated to observations. Typically $\delta_{\rm{ov}} = 0.12$ gives the best results \citep{starsdov}. It is calibrated for stars with initial masses between $2.5$ and $7\, \rm{M_\odot}$. 

For He-core burning models the inclusion of overshooting suppresses the growth of the semiconvective zone. This ultimately leads to the formation of a smaller combined convective and semiconvective region than if overshooting were not used. For the models with the modified Schwarzschild criterion, overshooting occurs where $\nabla_{\rm{r}} - \nabla_{\rm{a}} +0.15 >  - \delta$

By contrast, the stellar evolution code MESA defines semiconvective regions as those which are unstable to convection according to the Schwarzschild criterion but stable according to the Ledoux criterion \citep{mesa1, mesa2}. The MESA overshooting region is defined as $l_{\rm{ov}}H_P$ where $l_{\rm{ov}}$ is user defined and $H_P$ is the pressure scale height. Because $H_P \to \infty$ as $r \to 0$ the overshooting length can become very large for stars with small convective cores.

\subsection{Mixing Length Theory}
In the STARS code mixing length theory as described by \cite{mlt} is used. Near the core the pressure scale height $H_P \to \infty$. The mixing length $l$ is defined to be $\alpha H_P$ and sets the average distance travelled by convective elements. The mixing length parameter $\alpha$ is a user defined constant. Physically, convective elements cannot travel an infinite distance. As suggested by \cite{starssemiconv}, the mixing length is modified such that it cannot exceed the distance to the edge of the convective zone. This also has a modest effect on the mixing velocity $w$.

\section{Convective Tidal Dissipation}

\begin{figure}
	\includegraphics[width=\columnwidth]{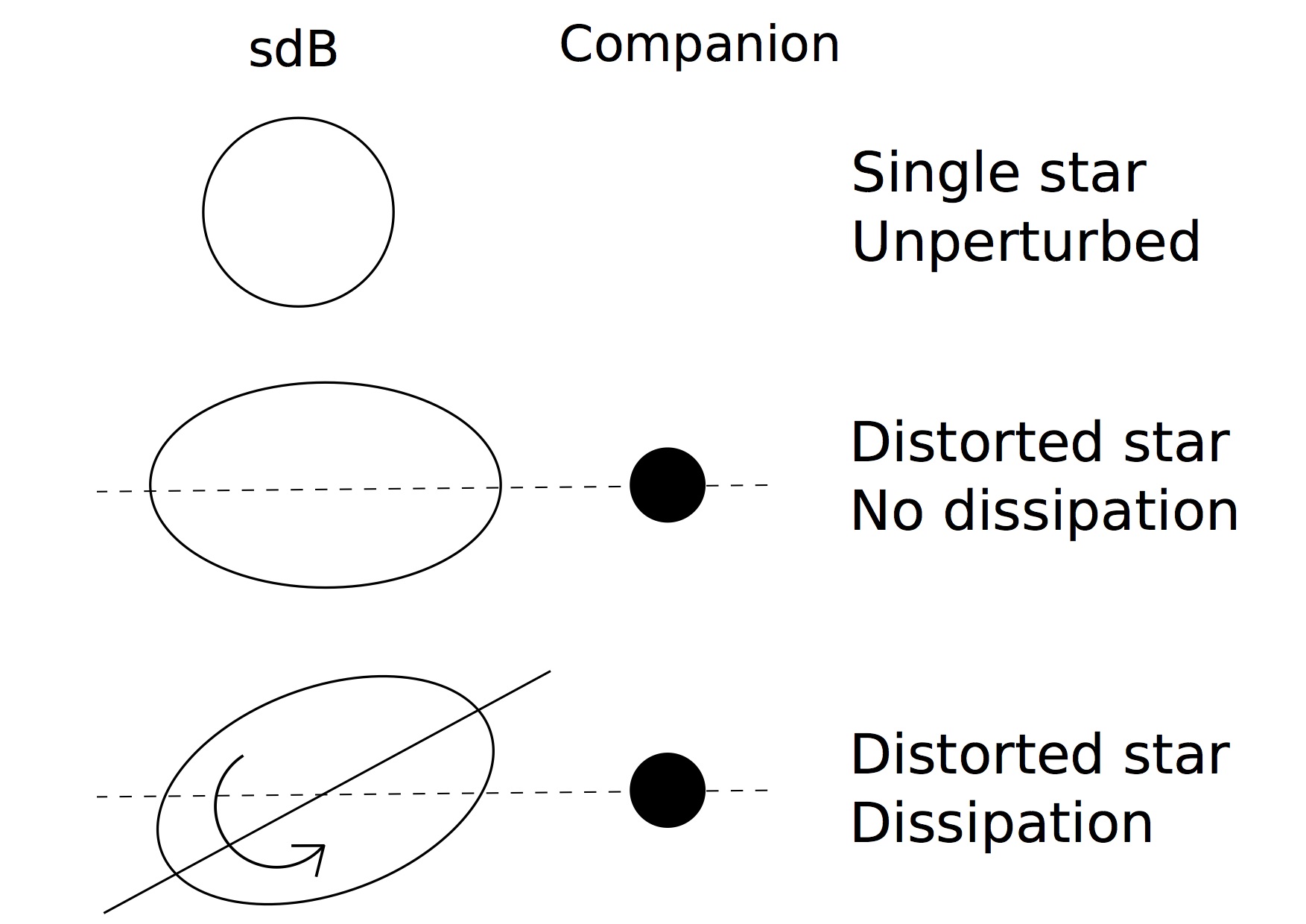}
	\caption{Schematic diagram of tidal interactions. The first panel shows a single, unperturbed star. The second panel shows a tidally distorted star which is either tidally synchronized or has no dissipation. The tidal bulge lies along the line connecting the centre of masses of the binary system. The final panel shows a non-synchronously rotating, tidally distorted star with a dissipation mechanism. The tidal bulge lags or leads the line connecting the centre of masses of the stars. This produces a torque which spins the star either up or down until it is rotating synchronously.}
	\label{fig:tidesdiag}
\end{figure}

The most efficient mechanism for tidal dissipation in sdB stars in close binaries is convective dissipation. Convection implies the bulk movement of material over large distances within the star. Turbulent viscosity in the convective region causes the tidal bulge to move away from the line connecting the centres of mass of the two stars. Fig. \ref{fig:tidesdiag} is a schematic diagram illustrating the dissipation.

The tidal synchronization time-scale $\tau_{\rm{sync}}$, owing to convective dissipation, as described by \cite{eggzbook} and \cite{eggstheory}, is 
\begin{equation}
    \tau_{\rm{sync}}\,=\,\frac{2}{9} \log\bigg(\frac{\omega-\Omega_0}{\omega-\Omega}\bigg) \bigg(\frac{M_1}{M_2}(1-Q) \bigg)^2 \bigg(\frac{I}{M_1R_1^2}\bigg)\frac{a^6}{R_1^6}\tau_{\rm{visc}},
    \label{eq:tsync}
\end{equation}
where $I$ is the moment of inertia of the star, $\tau_{\rm{visc}}$ is the viscous time, $Q$ is the dimensionless quadrupole moment, $a$ is the binary separation radius, $R_1$ is the radius of the dissipative region, $M_1$ is the mass of the dissipative region, $M_2$ is the mass of the companion, $\omega$ is the angular frequency of the binary, $\Omega_0$ is the initial spin angular frequency of the primary and $\Omega$ is the final spin angular frequency. The viscous time $\tau_{\rm{visc}}$ is 

\begin{equation}
    \tau_{\rm{visc}}\,=\,\frac{M_1R^2_1}{\int_0^{M_{1}}wl\gamma(r)\Psi(r) dm.},
    \label{eq:tvisc}
\end{equation}
where $\gamma(r)$ is a dimensionless structural property related to the coupling of the tides. The tides are described as fast when the orbital period is faster than the convective turnover time. In this circumstance the dissipation of the tides is damped in a way that depends on the turbulent spectrum of convective cells. 
\cite{zahn66} and \cite{tidedampchris} use a damping factor $\Psi (r)$ 
\begin{equation}
\Psi_1 (r)\,=\,\bigg|\frac{wP_{\rm{orb}}}{2l} \bigg|.
\end{equation}
\cite{damp2} have used
\begin{equation}
\Psi_2 (r)\,=\,\bigg|\frac{wP_{\rm{orb}}}{2l} \bigg|^2.
\end{equation}
\cite{hydrodis} revisited the problem with 3D hydrodynamical simulations and found better agreement between theory and observation with $\Psi_1(r)$. 

\subsection{The Tidal Synchronization Time-scale}

\begin{table*}
\caption{Convective mass, radius, dimensionless quadrupole moment and synchronization time for the models considered.}
\label{tab1}
\begin{tabular}{ |l|c|c|c|c|c|c| }
\hline
Model & $R_1/R_{\odot}$ & $M_1/M_{\odot}$ &$Q$& $(1-Q)^2$ & $\tau_{\rm{sync}}(\Psi_1)/\rm{Gyr}$ & $\tau_{\rm{sync}}(\Psi_2)/\rm{Gyr}$\\
\hline
$\rm{Early} \,\delta_{\rm{ov}}=0$ & 0.021 & 0.113 &0.01385 & 0.97250&36.47&588.63\\
$\rm{Late}\, \delta_{\rm{ov}}=0$  & 0.033 & 0.280 & 0.00890 & 0.98228&65.31&2448.11\\
$\rm{Early}\, \delta_{\rm{ov}}=0.12$ & 0.021 & 0.107 & 0.01389 & 0.97241&32.76&492.56\\
$\rm{Late}\, \delta_{\rm{ov}}=0.12$  & 0.024 & 0.151 & 0.00802 & 0.98402&62.78&1007.95\\
$\rm{Early} \, \Delta \nabla + 0.15$ & 0.042 & 0.317 & 0.01078 & 0.97856&74.02&4222.01\\
\rm{Late} \, $\Delta \nabla + 0.15$  & 0.053 & 0.441 & 0.00867 & 0.98274&12.52&60.24\\
\hline
\end{tabular}
\end{table*}

As can be seen in  Table 1, tidal synchronization time-scales for NY Vir predicted from standard models of sdB stars are close to or longer than the Hubble time. Tidal synchronization cannot occur before these models exhaust their core helium supplies, move off the EHB and on to a white dwarf cooling track. Because $\tau_{\rm{sync}}$ is inversely proportional to the radius of the dissipative region to the sixth power, simple calculations suggest that increasing the convective radius $r_{\rm{conv}}$ should substantially decrease the synchronization time. Somewhat surprisingly, increasing $r_{\rm{conv}}$ by a factor of 2.5 by modifying the Schwarzschild criterion only reduces the synchronization time by about an order of magnitude. The changes in the structural properties of the star affect the quadrupole tensor and so too the tides. Increases in the viscous time and mass of the convective region and decreases in the quadrupole moment and moment of inertia term counteract the effect of increasing the fractional convective radius.

The equation for tidal synchronization has multiple terms, all of which have an allowed physically constrained ranges. The dependence of the radial extent of the convective zone on the individual terms and their allowed ranges is now examined.  %The expression for $\tau_{\rm{visc}}$ can be substituted in to give
%\begin{equation}
%\frac{9M_2^2}{2a^6} \tau_{\rm{EHB}}> \log\bigg(\frac{\omega-\Omega_0}{\omega-\Omega}\bigg) \bigg(\frac{I}{M_1R_1^2}\bigg)\frac{M_1^3}{R_1^4}\frac{(1-Q)^2}{\int_0^{M_1}wl\gamma(r) dm}.
%\end{equation}

%\begin{figure}
%	\includegraphics[width=\columnwidth]{plots/tsyncrough.pdf}
%    \caption{The tidal synchronization time as a function of fractional convective radius. The blue points are created with no convective overshoot. The red points have no convective overshoot and a modified Schwarzschild criterion to create larger convective regions. The green points have an overshooting parameter of $\delta_{\rm{ov}}\,=\,0.12$. The vertical spread in the data is owing to use of different mixing length treatments. Increasing the convective radius by a factor of 2.8 only reduces the synchronization timescale by about an order of magnitude.}
%    \label{fig:tsync}
%\end{figure}

\subsection{The Dimensionless Quadrupole Moment}

The mass quadrupole tensor of an object describes the spatial distribution of the matter. If the object is a point source the quadrupole tensor vanishes. The dimensionless quadrupole moment $Q$ is given in Table \ref{tab1}. Varying $r_{\rm{conv}}/R_{\rm{sdB}}$ does not particularly change $Q$ because the early and late models with convective overshooting have the same fractional convective radius. However $Q$ is sensitive to the total radius and density of the star and $Q$ doesn't particularly change for the evolving models with no overshooting and a modified Schwarzschild criterion ($\Delta \nabla + 0.15$). For the models tested $(1-Q)^2$ is between 0.97 and 0.99. Because $(1-Q)^2$ is close to unity in all cases considered it does not have a substantial influence on the tidal synchronization time-scale. %To further examine the allowed values for the $(1-Q)^2$ term, $Q$ is calculated for the full grid of models used in \cite{preece} and plotted as a histogram.

%\begin{figure}
%	\includegraphics[width=\columnwidth]{plots/q.pdf}
%    \caption{The dimensionless quadrupole moment as a function of fractional convective radius of the sdB model.}
%    \label{fig:q}
%\end{figure}

\subsection{The Mass and Radius of the Convective Region}
\begin{figure}
	\includegraphics[width=\columnwidth]{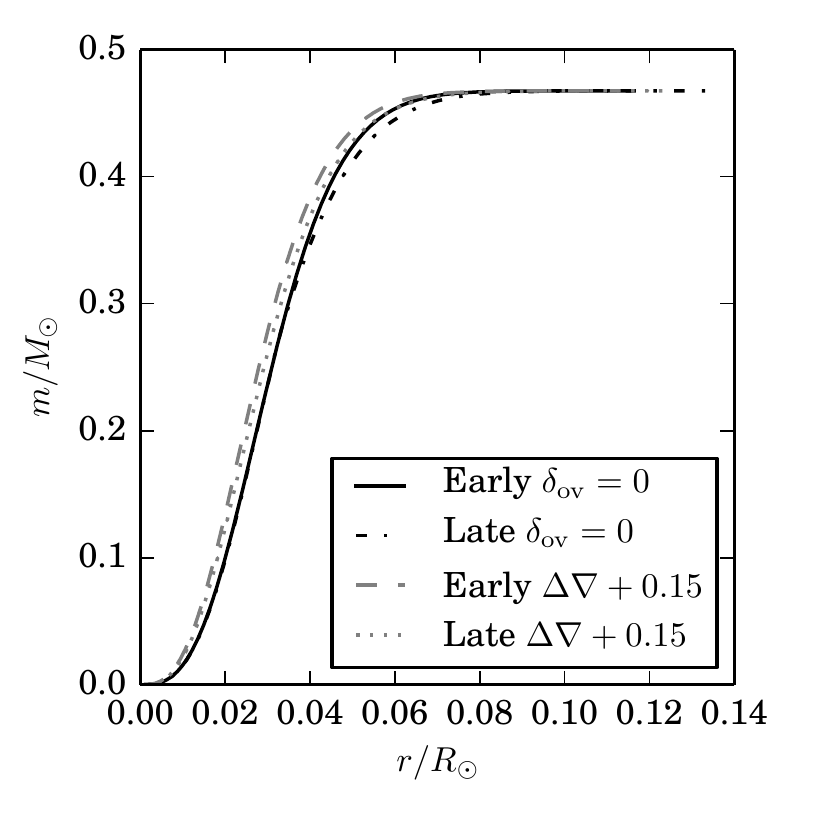}
    \caption{The mass $m$ contained within a sphere of radius $r$. The models with semiconvection and no convective overshooting are plotted in black and labelled $\delta_{\rm{ov}}=0$. The models with no convective overshooting had the same profiles and thus are not plotted. The modified Schwarzschild criterion models, labelled $\Delta \nabla + 0.15$ and plotted in grey, are denser.}
    \label{fig:mr}
\end{figure}

\begin{figure}
	\includegraphics[width=\columnwidth]{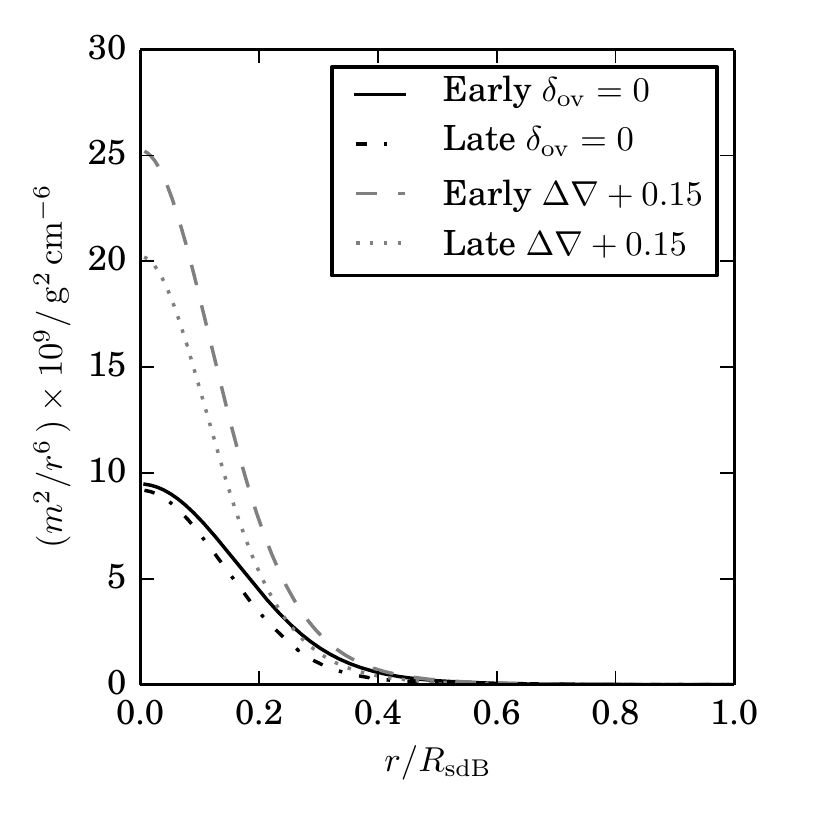}
    \caption{The mass to the second power over the radius to the sixth power as a function of fractional radius for the same models as in Fig. \ref{fig:mr}. }
    \label{fig:m2r6}
\end{figure}

The sdB star He cores are small but dense. The H-rich envelope is radially extended but accounts for a small amount of the mass. The mass as a function of radius can be seen in Fig.~\ref{fig:mr}. The outer regions of the star expand as the star evolves. In addition the high internal density means a small increase in the convective radius substantially increases the convective mass. When convective overshooting is used the radius of the convective region stays approximately the same but the mass increases by half. It is worth noting that whether a region is convective or radiative has little impact on the density profile. The models with the modified Schwarzschild criterion are denser than the standard models. Furthermore, the mass and radius term in Eq. \ref{eq:tsync} can be plotted as in Fig. \ref{fig:m2r6}. From this the mass and radius term can be constrained to be between $10^8$ and $7\times 10^9\,\rm{g^2\,cm^{-6}}$.

\subsection{Moment of Inertia Term}

\begin{figure}
	\includegraphics[width=\columnwidth]{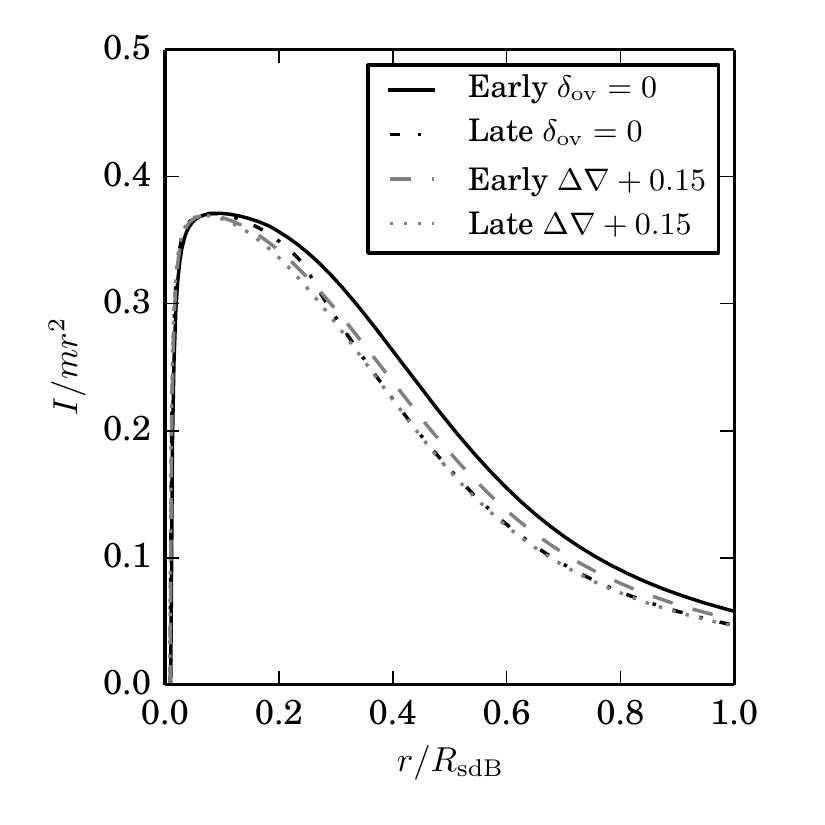}
    \caption{The ratio of the moment of inertia of the region enclosed to the moment of inertia if the matter were confined to a shell placed at the same radius as a function of fractional sdB radius.}
    \label{fig:krcc}
\end{figure}

For tidal calculations the ratio of the moment of inertia at the edge of the convective core to the moment of inertia if the mass were confided to a shell at the same radius is required. The overall dependence of the moment of inertia term on the fractional convective radius is displayed in Fig. \ref{fig:krcc}. This term lies between 0.15 and 0.37. 

\subsection{The Viscous Time}

Tidal interactions convert kinetic energy from tidal distortions into heat by dissipative processes whilst conserving angular momentum. This dissipation can be calculated from the square of the variation in the quadrupole tensor over time. As the companion moves around the sdB star the gravitational potential through the star changes cyclically. This changes the matter distribution and so affects the quadrupole tensor. If not synchronized, the tidal bulge moves around the star following the companion. This introduces a time dependent velocity field in the dissipative regions. The dissipation has a time-scale of $\tau_{\rm{visc}}$.  

\begin{figure}
	\includegraphics[width=\columnwidth]{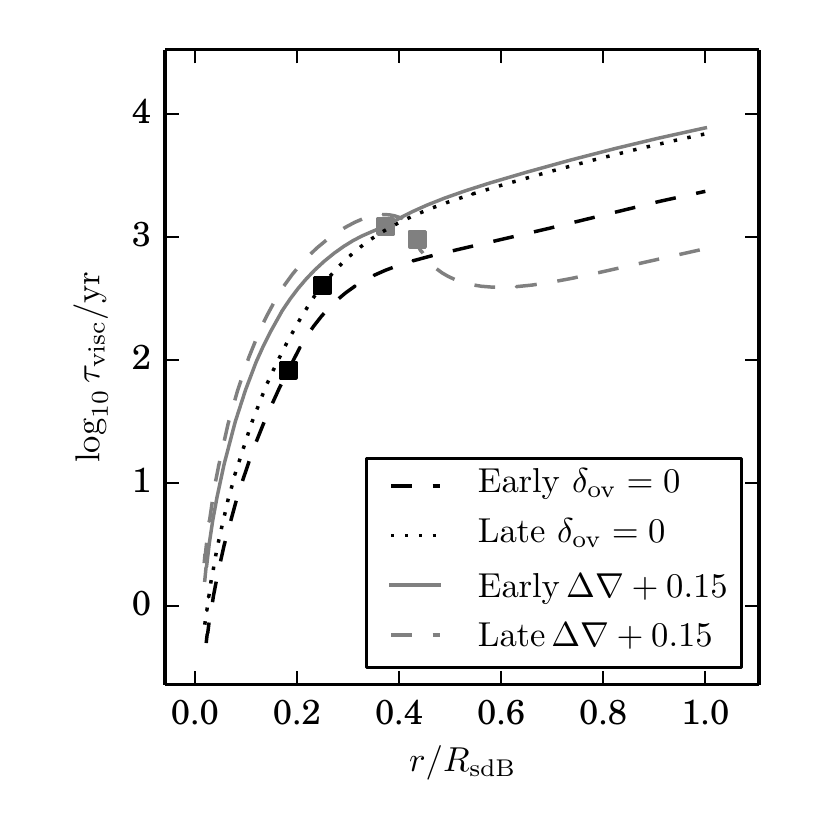}
    \caption{The viscous time-scale as a function of the convective radius as a fraction of the total radius. Each line represents a single evolutionary model. The points plotted are the viscous times predicted by the stellar models with each model's convective core radius. Increasing the radius of the convective region increases the viscous time-scale. A damping factor of $\Psi_1$ is used. }
    \label{fig:tviscallr}
\end{figure}

The viscous time as calculated by Eq. \ref{eq:tvisc} with $\Psi_1$ is shown in Fig. \ref{fig:tviscallr} for convective regions modified to extend throughout the star. The mixing length $l$ used is the distance to the edge of the convective region such that at $r=0$, $l=R_1$ and at $r = R_1$, $l=0$. The time-scales at the models' convective boundaries are also plotted. The $\delta_{\rm{ov}} = 0.12$ models are again omitted because they are almost identical to the $\delta_{\rm{ov}} = 0$ models. The mixing velocity $w$ is not well defined for regions which would be radiative. If $r<R_1$ we use $w$ from the models. If $r>R_1$, $w$ was given the same distribution but over the extended region. The dip in the Late $\Delta \nabla + 0.15$ models is due to a large peak in $\gamma(r)$ at the convective boundary. This is most likely an artifact of our modification to the Schwarzschild criterion. Without this peak the $\tau_{\rm{visc}}$ profile is almost identical to the Early $\Delta \nabla + 0.15$ profile. Overall, larger convective cores have longer viscous times.     

\subsection{Critical Viscous Time}
\begin{figure}
	\includegraphics[width=\columnwidth]{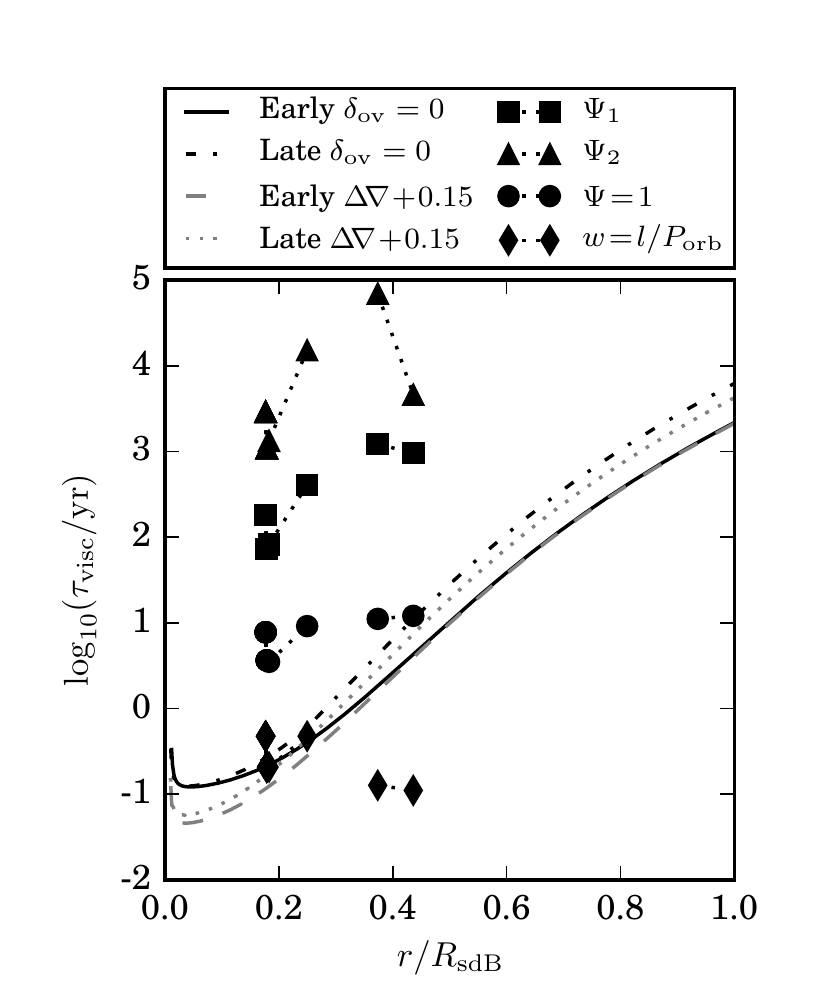}
    \caption{The viscous time-scale $\tau_{\rm{visc}}$ as a function of the fractional convective radius. The curves show the critical viscous time, for each convective radius, to achieve tidal synchronization during the EHB lifetime. Each line represents a single evolutionary model. The points plotted and connected with dotted lines are the viscous times predicted by the stellar models. The dotted lines connect the end points of evolutionary sequences with the same input physics ($\delta_{\rm{ov}}=0$, $\delta_{\rm{ov}}=0.12$ and $\Delta \nabla + 0.15$).  For completeness the evolutionary models with convective overshoot are also plotted. The early models without a modified Schwarzschild criterion both start at almost identical places. The spread in the points is due to different damping factors. The squares use $\Psi_1$, the triangles $\Psi_2$ and the circles $\Psi=1$. The diamonds have the convective velocity $w=l/P_{\rm{orb}}$, the minimum velocity required for the tides not to be considered fast. Increasing the fractional convective region increases the viscous time-scale.  }
    \label{fig:tvisc}
\end{figure}

All terms on the right hand side of Eq. \ref{eq:tsync} are known for any given mass and radius. The equation can be rearranged for the desired synchronization time. The maximum $\tau_{\rm{visc}}$ to synchronize the system in this time as a function of convective radius can be derived. The upper limit to the viscous time for synchronization within the EHB lifetime $\tau_{\rm{EHB}}\,=\,10^8\,\rm{yr}$ is plotted in Fig. \ref{fig:tvisc}. The viscous time as calculated with Eq. \ref{eq:tvisc} for each of the evolutionary models in Table \ref{tab1} is also plotted.

The $\tau_{\rm{visc}}$ calculated for the models with Eq. \ref{eq:tvisc} and the usual mixing theory estimates for the convective velocity are above the upper limits even when the damping factor is ignored. The $\tau_{\rm{visc}}$ increases as the convective core grows for all models with increasing radius owing to the high density of the material in the helium mantle. The damping factor is the most influential parameter. The choice of $\Psi$ stratifies $\tau_{\rm{visc}}$ by orders of magnitude. When the damping factor is excluded $\tau_{\rm{visc}}$ does not vary much between the models and is about $10\,\rm{yr}$. Fig. \ref{fig:tvisc} shows that doubling the radial extent of the convective region increases the viscous time-scale by approximately an order of magnitude. If the mixing velocity is increased such that $w=l/P_{\rm{orb}}$ the tides are no longer considered fast and so are not damped. The convective velocity is driven by the heat flux. Increasing the velocity would cause a substantial increase in the heat flux which would then change the temperature gradient and hence the structure in other significant ways. If the convective cells turnover without releasing all of their energy to the surroundings higher velocities can be reached without changing the overall heat flux. Some 3D hydrodynamical simulations of convective regions in stars have typical velocities which are much larger than those predicted by mixing length theory \citep{arnett3dhydro,avishaiv}.     

The derived upper limits are all within an order of magnitude of each other. If the radius of the dissipative region is small the viscous time must be less than a year for synchronization to be achieved. For convective regions which take up more than half of the sdB star by radius the critical viscous time is less than $10\,\rm{yr}$. This is more than the viscous time predicted when no damping is included.

%\begin{equation}
%\frac{9M_2^2}{2a^6}\tau_{\rm{EHB}} > \log\bigg(\frac{\omega-\Omega_0}{\omega-\Omega}\bigg) \bigg(\frac{I}{M_1R_1^2}\bigg)\frac{M_1^2}{R_1^6}(1-Q)^2\tau_{\rm{visc}},
%\end{equation}
%where $\tau_{\rm{EHB}}$ is the EHB lifetime and all the terms on the left hand side are known.

\section{Discussion}

%Typically, sdB stars spend around $10^8\,\rm{yr}$  on the EHB. For tidal synchronization to be a reasonable assumption to make for these stars they must synchronize fairly rapidly after arriving on the EHB, within say $10^7\,\rm{yr}$. Even in the most efficient cases the synchronization times for NY\,Vir are $10^9\,\rm{yr}$. The modifications made to the input physics at this stage are probably quite unphysical and reducing the estimates by a further two orders of magnitude would require further extreme alterations.

$J162256+473051$ is another HW\,Vir type system with a lower mass companion in a shorter-period orbit than NY\,Vir. Observations show that this star is rotating sub-synchronously. Calculations of the tidal synchronization time indicate that this system should synchronize more rapidly than NY\,Vir owing to its substantially smaller orbital separation. If $J162256+473051$ is neither expected nor observed to be synchronized, why should NY\,Vir appear to be synchronized?

HW\,Vir type systems are most likely formed via a common-envelope interaction \citep{han1}. The spin of the outer regions of the sdB star that subsequently forms are affected during this process. NY\,Vir has p-modes which propagate through the outer 55 per cent of the star. These p-modes are consistent with synchronization. The presence of p-modes means that the region must be radiative. However, synchronization could have been achieved during the common-envelope phase. NY\,Vir's companion is more massive and more radially extended than that of $J162256+473051$ so its companion should have had more of an effect during this phase.

%\begin{figure}
%	\includegraphics[width=\columnwidth]{plots/ttide.pdf}
%    \caption{Plot of ttide}
%    \label{fig:obs}
%\end{figure}

\section{Conclusions}

Asteroseismic evidence for rotational and orbital synchronization in the hot subdwarf binary NY\,Vir is at variance with our previous theoretical predictions of tidal synchronisation in such stars. 
Because the tidal synchronization time-scale is inversely proportional to the radius of the convective region to the sixth power artificial extensions to the convection boundary have been examined to see whether a larger convective region could account for the observed  synchronization. 
Increasing the radius of the convective region by a factor of 2.5 decreases tidal synchronization times by less than one order of magnitude, insufficient to bring NY\,Vir close to synchronization within its core-helium burning (or extended horizontal-branch) lifetime. 

The individual terms of Eq. \ref{eq:tsync} were examined to test how much each contributes to the synchronization time and what constraints may be placed on the quantities contained in them. 
The boundary of the convective core was moved outwards to see if there was a radius at which tidal synchronization could occur without any modifications to the theory. 
It was found that even making the stars fully convective would not be sufficient because the orbital periods are shorter than the convective turnover time and consequently dissipation of the tides is damped. 
The damping factor and choice of mixing length theory are the areas of largest uncertainty. 
If the convective mixing velocity is increased such that $w>l/P_{\rm{orb}}$ all the models predict tidal synchronization within the EHB lifetime. Some 3D hydrodynamical simulations of convective regions predict velocities substantially larger than those calculated with MLT. If these calculations prove correct, tidal synchronization might result from invocation of non-classical convection physics. 

In seeking an alternative explanation for the synchronization of NY\,Vir, we note that the common-envelope phase is not well understood. 
If the tides do not synchronize on the EHB it is possible that at least the outer layers of the sdB star were synchronized during the common-envelope phase.

\section*{Acknowledgements}

Research at the Armagh Observatory and Planetarium is supported by a grant-in-aid from the Northern Ireland Department for Communities. HPP acknowledges support from Nicky Saltiel, Churchill College and the UK Science and Technology Facilities Council (STFC) Grant No. ST/M502268/1. CSJ acknowledges support from STFC Grant No. ST/M000834/1. CAT thanks Churchill College for his fellowship.

%%%%%%%%%%%%%%%%%%%%%%%%%%%%%%%%%%%%%%%%%%%%%%%%%%



%\bibliographystyle{mnras}

\bibliographystyle{mnras}
\bibliography{references}

%%%%%%%%%%%%%%%%%%%%%%%%%%%%%%%%%%%%%%%%%%%%%%%%%%

%%%%%%%%%%%%%%%%% APPENDICES %%%%%%%%%%%%%%%%%%%%%

%\appendix


%%%%%%%%%%%%%%%%%%%%%%%%%%%%%%%%%%%%%%%%%%%%%%%%%%

% Don't change these lines
\bsp	% typesetting comment
\label{lastpage}
\end{document}